\documentclass[final,5p,times]{elsarticle}
\usepackage{lineno}

\usepackage{graphicx}
\usepackage{amssymb}
\usepackage{siunitx}
\usepackage{pdfpages}
\usepackage{subfigure}
\usepackage{float}
\usepackage{amsmath}
\usepackage{color}
\usepackage{ulem}

\usepackage{xspace}
\usepackage{booktabs}
\usepackage{multirow}

\newcommand{\figref}[1]{Fig.~\ref{#1}}
\newcommand{\Figref}[1]{Figure~\ref{#1}}	
\newcommand{\secref}[1]{Sec.~\ref{#1}}
\newcommand{\Secref}[1]{Section~\ref{#1}}

\newcommand{\tabref}[1]{Tab.~\ref{#1}}
\newcommand{\Tabref}[1]{Table~\ref{#1}}

\newcommand{\NeCOtwoNtwo}{Ne-CO$_2$-N$_2$ (90-10-5)\xspace}
\newcommand{\NeCOtwo}{Ne-CO$_2$ (90-10)\xspace}

\newcommand{\ArCOtwo}{Ar-CO$_2$ (90-10)\xspace}
\newcommand{\ArCOtwoThirty}{Ar-CO$_2$ (70-30)\xspace}

\newcommand{\tInt}{\ensuremath{t_{\mathrm{int}}}\xspace}
\newcommand{\Qcrit}{\ensuremath{Q_{\mathrm{crit}}}\xspace}
\newcommand{\chiSquared}{\ensuremath{\chi^{2}}\xspace}

\begin{document}

\begin{frontmatter}
\title{
Charge density as a driving factor of discharge formation in GEM-based detectors}
\address[exc]{Excellence Cluster Universe, Technische Universit\"at M\"unchen, Boltzmannstr. 2, 85748 Garching, Germany}
\address[tum]{Physik Department E62, Technische Universit\"{a}t M\"{u}nchen, James-Franck-Str. 1, 85748 Garching, Germany}
\address[bic]{Dipartimento di Fisica G.Occhialini - Universit\'{a} Milano-Bicocca, Piazza della Scienza 3, 20126 Milano, Italy}

\author[exc,tum]{P.~Gasik\corref{cor1}} \ead{p.gasik@tum.de}
\author[exc,tum]{A.~Mathis\corref{cor1}} \ead{andreas.mathis@ph.tum.de}
\author[exc,tum]{L.~Fabbietti} 
\author[exc,bic]{J.~Margutti\corref{cor2}} 

\cortext[cor1]{Corresponding author}

\cortext[cor2]{Currently at Utrecht University, NIKHEF}

\begin{abstract}

We report on discharge probability studies with a single Gas Electron Multiplier (GEM) under irradiation with alpha particles in Ar- and Ne-based gas mixtures. The discharge probability as a function of the GEM absolute gain is measured for various distances between an alpha source and the GEM. We observe that the discharge probability is the highest when the charge deposit occurs in the closest vicinity of the GEM holes, and that the breakdown limit is lower for argon mixtures than for neon mixtures.

Our experimental findings are in line with the well-grounded hypothesis of the charge density being the limiting factor of GEM stability against discharges.
A detailed comparison of the measurements with GEANT4 simulations allowed us to extract the critical charge density leading to the formation of a spark in a GEM hole. 
This number is found to be within the range of $\SIrange[range-units = brackets]{5}{9}{}\times10^6$ electrons after amplification, and it depends on the gas mixture.

\end{abstract}


\end{frontmatter}

\section{Introduction}

The requirements of a new generation of experiments in particle physics determine the development of new detectors. Novel devices must handle the envisaged high luminosities as well as a substantial increase in active detector area. Among the innovative detector techniques, Micro-Pattern Gaseous Detectors (MPGD), and in particular the Gas Electron Multiplier (GEM) \cite{GEM_Sauli}, have become a widely used technology for high-rate experiments like COMPASS \cite{COMPASS, COMPASS2}, LHCb \cite{LHCb} or TOTEM \cite{TOTEM, TOTEM2} and are foreseen
for new experiments or upcoming upgrades, such as sPHENIX \cite{sPHENIX} or the upgraded ALICE TPC \cite{TPCU_TDR, TPCU2}.
The key aspects for long-term operation of such detectors in the harsh environment of high-rate experiments are radiation hardness, ageing resistance and stability against electrical discharges. In particular, the latter may pose a threat to the integrity of the GEM foil, as they may cause irreversible damages to the detector, ranging from enhanced leakage currents to permanent electric short circuits that render the detector effectively blind.

In studies with parallel-plate counters \cite{Raether}, breakdown has been measured to occur when the total charge in the a\-va\-lanche exceeds a critical value of the order of $10^8$ electron--ion pairs (the so-called Raether limit). Correspondingly, an avalanche size of $\mathcal{O}(10^7)$ electron-ion pairs is expected to be the limiting factor of the micro-pattern gaseous detectors stability against electrical discharges \cite{Bressan}.

Typical high-rate capabilities considered for GEM-based \linebreak trackers are of the order of \SI{1}{\mega\hertz/\centi\meter\squared} \cite{COMPASS, COMPASS2, LHCb, TOTEM2}. Taking into account the geometry of such detectors (few mm drift gaps) and gas mixtures (Ar-based), a single MIP-like particle impinging on the detector plane creates an average primary charge density of a few electrons per cm$^2$. Even at these rates, taking into account the electron drift time, no pile-up is expected. The resulting primary charge density is far below the stability limits, even if multiplied by typical gain values of $\mathcal{O}(10^4)$ at which such detectors are operated. However, the occurrence of high charge densities, possibly produced by the presence of highly ionising particles in the closest vicinity of the GEM foil, may significantly alter the stability of the detector. Indeed, an alpha particle may liberate up to 10$^4$ more electrons per cm than a MIP.

So far, the only comprehensive discharge studies with single and multiple GEM detectors exposed to highly-ionising alpha particles were reported in \cite{Bachmann} and concern mainly Ar-based gas mixtures. By sharing the amplification process between two or more successive GEMs, the maximum sustainable gain is shifted upward by several orders of magnitude. The charge spread during the transport between adjacent foils dilutes the charge densities approaching the last amplification stage.

There are several ways to distribute the voltages among all electrodes in a multi-GEM set-up. According to \cite{Bachmann}, where the detector stability was studied using an external alpha source, the most stable HV configuration features an asymmetric choice of potential differences across the GEM electrodes, in which the highest amplification occurs in the first foil in the stack and successively decreases towards the readout anode. Such a configuration has been successfully used in the GEM-based trackers of COMPASS \cite{COMPASS, COMPASS2}.

The experience of the LHCb experiment at CERN-LHC with their triple GEM detectors points towards a reduction of the gain by lowering the voltages applied to the first two amplification stages to the level of the third GEM in a stack \cite{LHCb2}. Such a symmetric distribution helps to eliminate the occurrence of fatal discharges in the first GEM of a stack due to (what is believed to be) neutrons converting in the detector material.

In case the application demands for a different configuration of the potentials applied to the electrodes, the stability of the detector may become compromised. For the application of GEMs in a TPC, the suppression of the ion back-flow is crucial in order to keep the distortions of the drift field by the back drifting ions at a tolerable level (see e.g.~\cite{ALICE_IBF} and references therein). A typical suppression of the ion feedback is conducted by subsequently increasing the gain of each GEM foil in the stack towards the readout anode, as in this scenario ions created in the inner layers are blocked more efficiently. The strong amplification in the last stage, in combination with the pre-amplified charges from the upper GEMs may enhance the probability of reaching the critical amount of charge and thus significantly alter the stability with respect to electrical discharges. 

In all scenarios, the overall stability is driven by the probability of encountering a discharge in any of the GEMs. This clearly demand for further studies of the intrinsic stability of GEM-based detectors. Of particular interest are therefore measurements of the breakdown limits which can be obtained in a single-GEM detector, which will allow for the extrapolation of the results to the situation in an arbitrary stack operated under any HV settings. By measuring the stability of the simplest configuration, including only one GEM foil, the intrinsic stability of the GEM foil against electrical discharges can be extracted independently from effects as the presence of transfer fields or charge sharing and spreading between the foils.

In this paper we intend to determine critical charge limits for a Gas Electron Multiplier by irradiating a single-GEM detector with alpha particles.
The studies are performed in Ne- and Ar-based gas mixtures to investigate the influence of basic gas properties on the detector stability.

The paper is structured as follows. \Secref{sec:setup} introduces the experimental setup and the main properties of the employed gas mixtures. The measurements and their interpretation are described in \secref{sec:results}. \Secref{sec:sim} describes the simulation framework and the corresponding analysis procedure which allow to extract final values of the critical charge density. \Secref{sec:summary} concludes the paper with a summary of the main findings.
\section{Experimental setup}
\label{sec:setup}

\subsection{GEM detector}
\begin{figure}[h]
  \centering
   \includegraphics[width=\linewidth]{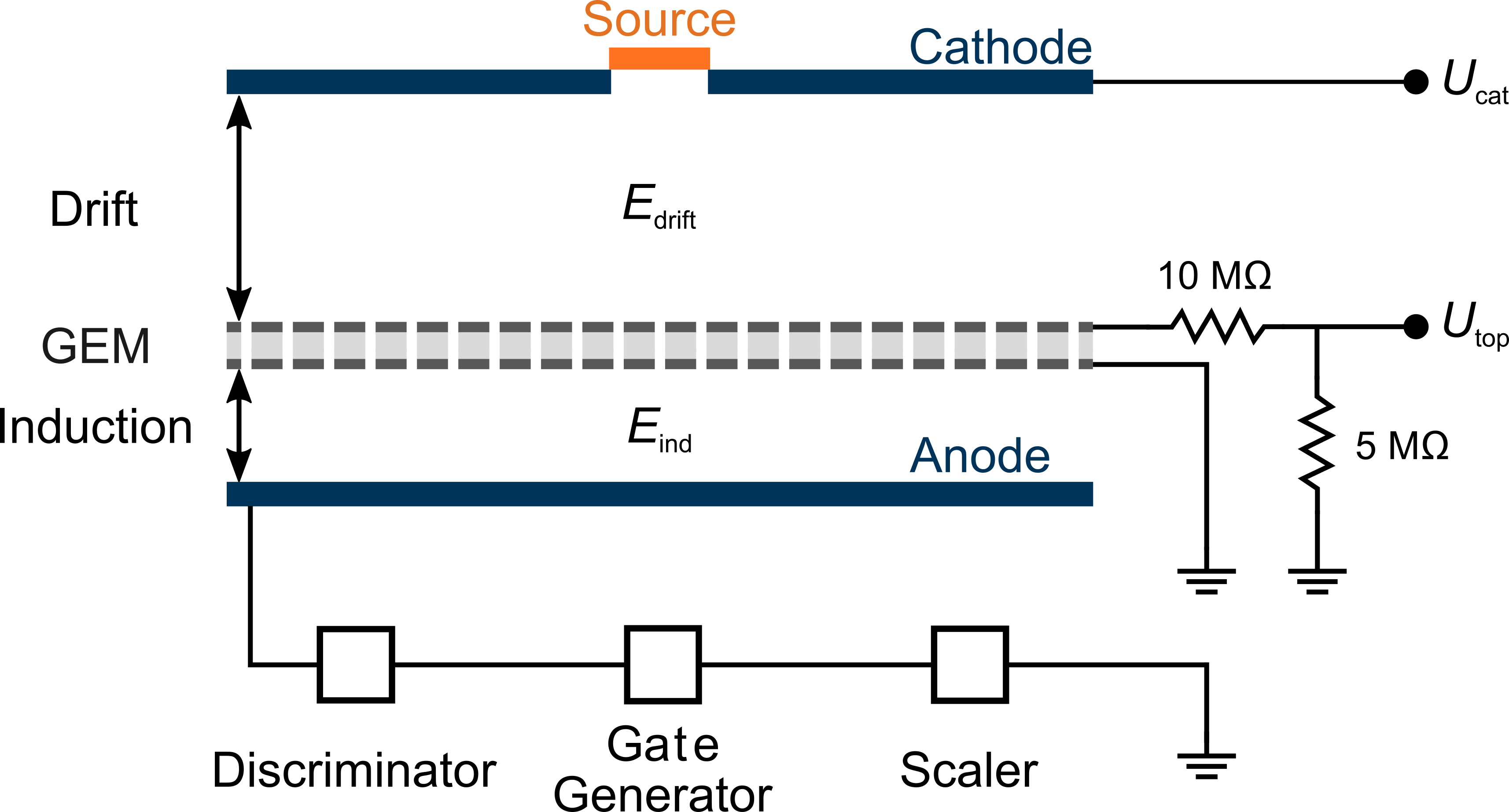}
  \caption[Schematic picture of the setup used for GEM discharge probability studies]{(\textit{Colour online}) Schematic picture of the single-GEM setup used for discharge probability studies.}
  \label{fig:sparksetup}
\end{figure}

A prototype was prepared to study the stability of single GEM foils against electrical discharges. \Figref{fig:sparksetup} shows a sche\-ma\-tic drawing and the powering scheme of this setup. 

The detector vessel contains a 10$\times$\SI{10}{\centi\meter\squared} GEM holder, a drift cathode and a readout anode. Both the cathode and the anode are made of a \SI{1.5}{\milli\meter} thick PCB coated on one side with copper (${\sim}10\times$\SI{10}{\centi\meter\squared}$)$. In the middle of the cathode plate, a \SI{7.8}{\milli\meter} diameter hole allows to irradiate the GEM plane with a radioactive source. A single GEM foil is installed \SI{2}{\milli\meter} above the readout plane. The GEM foil, produced at the CERN PCB workshop with the double-mask technology, is of standard design: \SI{50}{\micro\meter} thick polyimide (Apical) covered on both sides with \SI{5}{\micro\meter} layers of copper, with \SI{50}{\micro\meter} (\SI{70}{\micro\meter}) inner (outer) hole diameter and a pitch of \SI{140}{\micro\meter}. The drift gap length (distance between the cathode and the GEM stack) can be adjusted between 10 and 77\,mm.

Two voltages are applied to the detector: a cathode potential $U_{\mathrm{cat}}$ and, through a 10\,M$\Omega$ protection resistor, a GEM potential $U_{\mathrm{top}}$. Both voltages are applied using two independent channels of an ISEG EHS F060n \SI{6}{\kilo\volt} unit. A \SI{5}{\mega\ohm} resistor to ground for the top GEM electrode is chosen to ensure a safe and fast discharge of the GEM after a power supply trip.
The bottom side of the GEM is kept at ground potential.

$U_{\mathrm{cat}}$ and $U_{\mathrm{top}}$ define the drift field $E_{\mathrm{drift}}$ above the GEM and the potential difference across the GEM foil $\Delta U_{\mathrm{GEM}}$ (see \figref{fig:sparksetup}). A drift field of \SI{400}{\volt/\centi\meter}, used for all measurements presented in this work, matches the value of the drift field planned for the upgraded ALICE TPC~\cite{TPCU_TDR,TPCU2}. The choice of $E_{\mathrm{drift}}$ defines the drift properties in a given gas, which are discussed in Sec.~\ref{sec:gas}.
The value of the induction field ($E_{\mathrm{ind}}$) between the bottom side of the GEM and the grounded readout anode is set equal to zero in order to measure the absolute gain of the GEM foil (see \secref{sec:gain} for more details) and to decouple the effect of the induction field on the discharges at the GEM.

\begin{figure}[h]
  \centering
   \includegraphics[width=\linewidth]{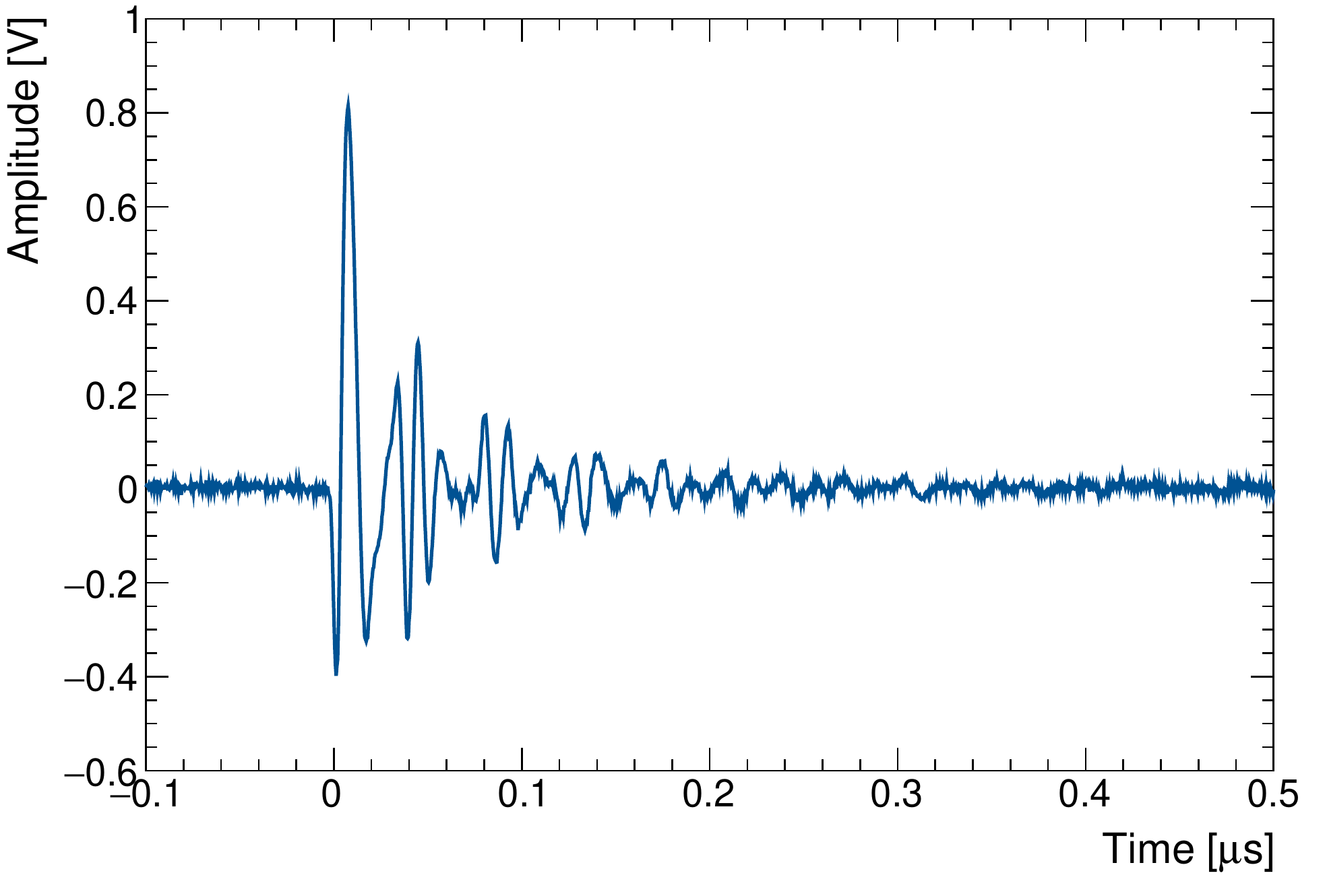}
  \caption[A typical signal associated with a spark in a small prototype]{A typical scope signal associated with a spark in the single-GEM setup, attenuated by 30\,dB.}
  \label{fig:smallspark}
\end{figure}

The occurrence of a spark in the GEM foil is detected employing the readout scheme shown in \figref{fig:sparksetup}. 
The signal induced on the pad plane is processed by a discriminator unit. The threshold of the discriminator is set to a few hundreds of a \si{\milli\volt} to filter out signals induced by alpha particles and to trigger on discharge signals with a much higher amplitude (a typical discharge signal is displayed in \figref{fig:smallspark}). 

The signal waveform shows modulations which are due to the inductance of the system, resulting in periodic oscillations.  Therefore, a gate of $\mathcal{O}($\SI{100}{\micro\second}$)$ is opened when the discriminator threshold is exceeded. A scaler registers the resulting counts. 
It should be noted that with this procedure no dead-time is introduced to the measurement (see \secref{sec:results:gas} for details).
\subsection{Radiation source}
\label{sec:spark:sources}
For the reported discharge studies, an alpha-emitting mixed nuclide source is used which consists of $^{239}$Pu, $^{241}$Am and $^{244}$Cm. The energies of the respective alpha particles emitted by the different constituents are summarised in \tabref{tab:Source}.

\begin{table}[h]
\centering
\caption{Parameters of the mixed-nuclid source. The relative intensity of the most important decay channels is given per parent nucleus \cite{EckertuZiegler}.}
\begin{tabular}{l c c}            
\toprule
\multirow{2}{*}{\textbf{Parent nucleus}} & \textbf{Energy} & \textbf{Relative intensity} \\
{}    & \textbf{[\si{\mega\electronvolt}]} & \textbf{[\si{\percent}]} \\
\midrule
\multirow{3}{*}{$^{239}$Pu} & 5.105 & 11.5 \\
							& 5.143 & 15.1 \\
							& 5.155 & 73.4 \\
							\hline
\multirow{3}{*}{$^{241}$Am} & 5.388 & 1.4 \\
                            & 5.442 & 12.8 \\
                            & 5.486 & 85.2 \\
\midrule
\multirow{2}{*}{$^{244}$Cm} & 5.763 & 23.3 \\
							& 5.805 & 76.7 \\
\bottomrule
\end{tabular}
\label{tab:Source}
\end{table}

The active material of the source is deposited on a coin-like area in a form of an 8 mm diameter circle. To determine the rate of the source, an ADC spectrum of alpha particles is recorded within a given time and the background is subtracted. The remaining counts determine the rate $R$, which is $R\approx600$\,Hz. The rate is measured after each positioning of the source on the cathode or each modification of the drift gap length. 

The discharge probability is defined by the ratio of the discharge rate to the measured source rate:
\begin{equation}
P = \frac{N}{tR},
\end{equation}
where $N/t$ is the discharge rate measured by the number of sparks $N$ recorded within the measurement time $t$, and $R$ is the rate of the source. 

\subsection{Gas mixtures}
\label{sec:gas}
Four different gas mixtures are used in the measurements presented in this work: Ar-CO$_2$ (70-30) as reference to previous works \cite{Bachmann}, Ar-CO$_2$ (90-10), Ne-CO$_2$ (90-10) and the ALICE TPC mixture Ne-CO$_2$-N$_2$ (90-10-5)\footnote{The gas ratio should be normalised to the sum of all components, which corresponds to (85.7-9.5-4.8) in standard notation.} \cite{JINST-gasik}. \Tabref{tab:sim:gas} summarises the basic properties of the gas mixtures. 
\begin{table*}[ht]
\centering
\caption{Properties of the gas mixtures employed in this study. The electron drift velocity and diffusion coefficients are evaluated under Normal Temperature and Pressure (NTP) conditions at a nominal drift field of \SI{400}{\volt / \centi\meter} and in absence of a magnetic field. The last column displays the approximate range of particles emitted by the mixed alpha source in the respective gas obtained with GEANT4 simulations. See text for details.}
\begin{tabular}{l *5c}
\toprule

\multirow{2}{*}{\textbf{Gas}}  & \textbf{\textit{v}}$_{\mathrm{\textbf{d}}}$ & \textbf{\textit{D}}$_{\mathrm{\textbf{L}}}$ & \textbf{\textit{D}}$_{\mathrm{\textbf{T}}}$ & \textbf{\textit{W}}$_{\mathrm{\textbf{i}}}$  & \textbf{\textit{r}}$\mathbf{_{\alpha}}$ \\
 & \textbf{[\si{\centi\meter / \micro\second}]} & \textbf{[\si[parse-numbers=false]{\sqrt{\centi\meter}}]} & \textbf{[\si[parse-numbers=false]{\sqrt{\centi\meter}}]} & \textbf{[\si{\electronvolt}]} & \textbf{[\si{\centi\meter}]} \\
\midrule
\ArCOtwoThirty & 0.932 & 0.0138 & 0.0145 & 28.1 & 4.2\\
\ArCOtwo & 3.26 & 0.0244 & 0.0268 & 28.8 &  4.8 \\
\NeCOtwo & 2.66 & 0.0223 & 0.0219 & 38.1 & 6.8 \\
\NeCOtwoNtwo & 2.52 & 0.0218 & 0.0224 & 37.3 & 6.9 \\
\bottomrule
\end{tabular}
\label{tab:sim:gas}
\end{table*}
Electron drift velocity ($v_{\mathrm{d}}$), longitudinal ($D_{\mathrm{L}}$) and transverse ($D_{\mathrm{T}}$) diffusion coefficients, and effective ionisation potentials ($W_{\mathrm{i}}$) are evaluated using Magboltz \cite{Garfield}, whereas the approximate value of the maximum range ($r_{\alpha}$) of an alpha particle from the mixed radiation source in a given gas mixture was extracted using GEANT simulations (see \secref{sec:geant} for more details). 

 \subsection{Absolute gain determination}
 \label{sec:gain}
All discharge probability measurements presented here have been performed as a function of the GEM absolute gain $G_{\mathrm{abs}}$. 
For the characterisation of a GEM foil, the difference is made between the GEM absolute gain and the so-called effective gain  $G_{\mathrm{eff}}=\epsilon_{\mathrm{coll}}\times G_{\mathrm{abs}}\times\epsilon_{\mathrm{extr}}$, where $\epsilon_{\mathrm{coll}}$ and $\epsilon_{\mathrm{extr}}$ are the electron collection and extraction efficiencies, respectively. 
According to our FEM calculations using the COMSOL Multiphysics$\textsuperscript{\textregistered}$ software \cite{COMSOL} the drift field of \SI{400}{\volt/\centi\meter} used in all the measurements assures \SI{100}{\percent} collection efficiency for primary electrons reaching the GEM foil. 
In order to allow for a proper interpretation of the measurements, the latter are conducted at zero induction field in order to avoid any bias introduced by a possible focussing effect which effectively enhances the local charge density inside the hole and accordingly increases the discharge probability.
With $E_{\mathrm{ind}}=0$ the extraction efficiency to the pad plane $\epsilon_{\mathrm{extr}}=0$, which means that all electrons produced in the avalanche process are collected at the bottom side of the GEM foil. Measuring the amplification current at this electrode, however, corresponds to a situation in which all electrons are read out and we thus find $G_{\mathrm{abs}} = G_{\mathrm{eff}}$.
The absolute gain of the GEM is evaluated as the ratio of the amplification current measured at the bottom electrode of the GEM to the primary ionisation current, created by the ionisation in the drift volume of the detector. The latter is measured at the top GEM electrode (keeping $\Delta U_{\mathrm{GEM}}=0$ and the bottom GEM electrode  grounded) at the nominal drift field. The radiation source used in these studies (see \secref{sec:spark:sources}) generates primary currents in the range of 10 to 20\,pA. The measurement is performed using a Keithley Instruments Model 6517B Electrometer in a low-noise environment provided by a shielding box surrounding the experimental setup. The final value of primary current is corrected for residual currents (offset) measured before applying potential to the electrodes.

\begin{figure}[h]
  \centering
     \includegraphics[width=\linewidth]{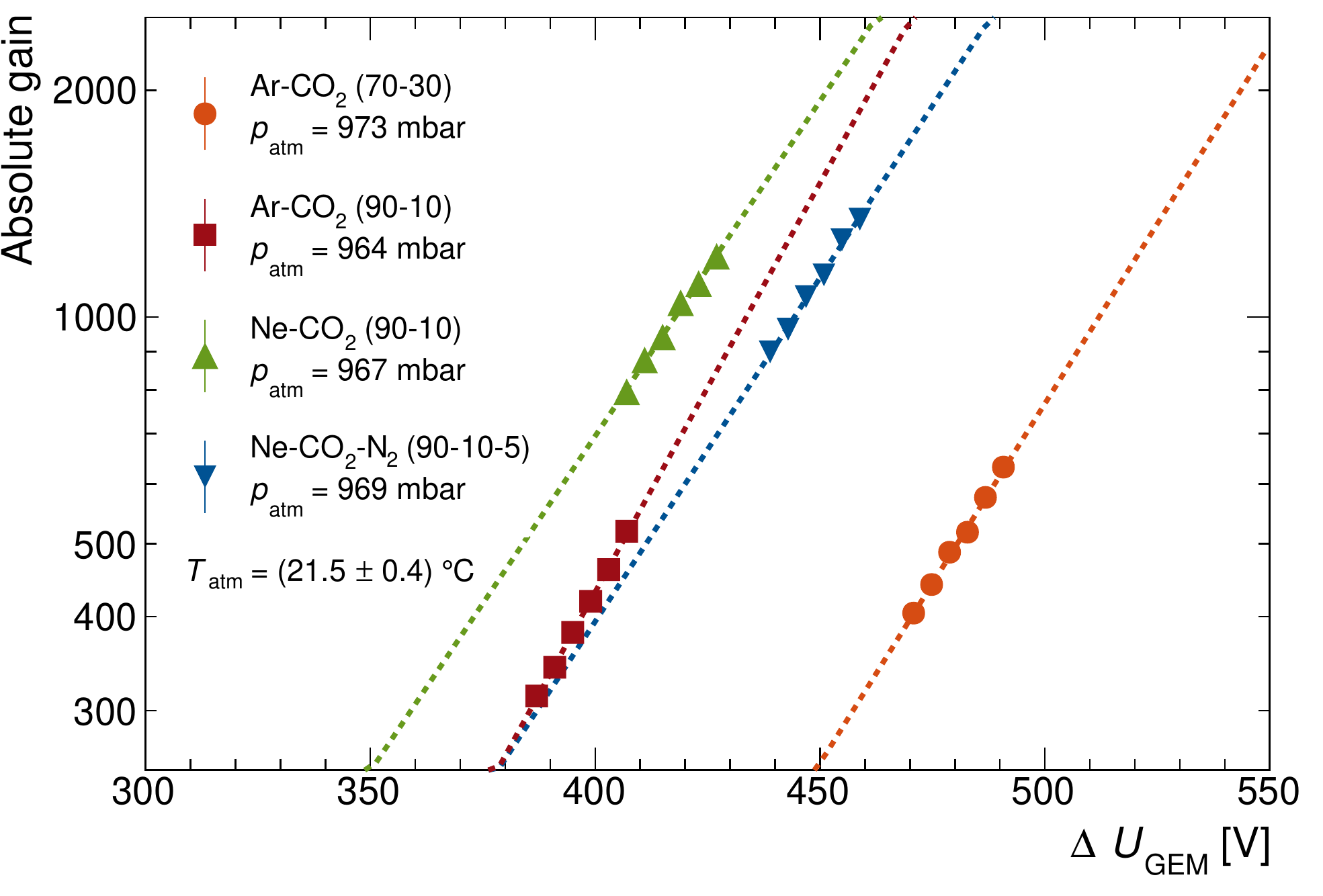}
  \caption[The GEM absolute gain as a function of the potential difference across the GEM foil.]{(\textit{Colour online}) The GEM absolute gain as a function of the potential difference across the GEM foil $\Delta U_{\mathrm{GEM}}$ for all gas mixtures employed in the course of our studies. The error bars are the same size or smaller than the symbols, while the dashed lines represent exponential fits to the data points. Atmospheric pressure ($p_{\mathrm{atm}}$) and temperature ($T_{\mathrm{atm}}$) are indicated in the legend.}
  \label{fig:1gemgain}
\end{figure}

\Figref{fig:1gemgain} shows typical GEM absolute gain values measured as a function of the potential difference across the foil for the four gas mixtures. It should be noted that the values of the absolute gain are very high for single GEMs as the detector was operated at high voltages in order to induce measurable discharge rates. We consider the gain to follow an exponential behaviour in the limited range of interest. Therefore, the gain curves are fitted with an exponential function in order to extrapolate or interpolate the absolute gain value to the region where no data points are available.

\section{Discharge probability in a single GEM}
\label{sec:results}

\subsection{Gas mixture dependence}
\label{sec:results:gas}
\begin{figure*}[p]
\centering
\includegraphics[width=\textwidth]{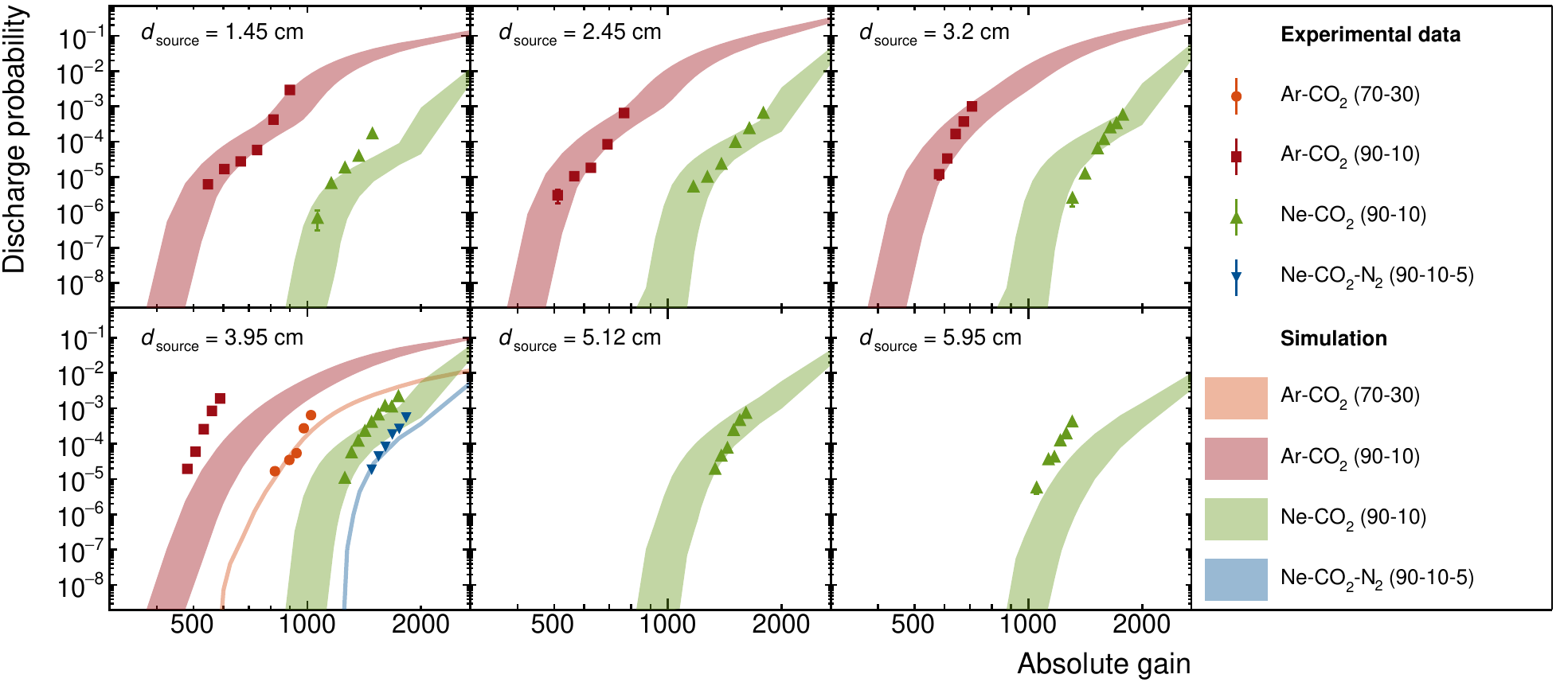}
\caption[Discharge probability as a function of the GEM absolute gain.]{(\textit{Colour online}) Discharge probability as a function of the GEM absolute gain. The bands indicate the outcome of the simulation, while the points correspond to measurements. The integration time in the simulation is \SI{50}{\nano\second} for \NeCOtwo, \SI{30}{\nano\second} for \ArCOtwo and \SI{40}{\nano\second} for \ArCOtwoThirty and \NeCOtwoNtwo. The uncertainties of the measurement are typically smaller than the marker size. The width of the simulation bands is related to the range of the value of critical charge density. See text for details.}
\label{fig:sim:Multiplication}
\end{figure*}

\Figref{fig:sim:Multiplication} shows the discharge probability measured as a function of the GEM absolute gain for different distances $d_{\mathrm{source}}$ between the source and the GEM foil. 

The values of discharge probability spread over several orders of magnitude for measurements at a given distance. With the  source rate of ${\sim}600$\,Hz, the measurement of the probability values higher than $10^{-3}$ is experimentally difficult as the corresponding discharge rate is higher than 1\,Hz. The average time between discharges is thus close to the system dead-time of ${\sim}$\SI{200}{\milli\second}, which is the time needed after a discharge to restore the nominal value of $\Delta U_{\mathrm{GEM}}$ across the GEM. The measured value of discharge probability may thus be underestimated, as the detector is not operated at full voltage. To correct for this effect we assume the spark occurrence undergoes a Poisson distribution and, for the measured spark rate, calculate the probability to encounter a discharge within a time interval corresponding to the detector's dead-time. For discharge rates of 1\,Hz the resulting probability is ${\sim}$\SI{16}{\percent} (at the level of the statistical error) which is also the measure of how much the final discharge probability is underestimated. For lower discharge rates the corrections are negligible. In order to avoid major corrections to the high-rate measurements, we do not continue measurements for spark rates larger than ${\sim}1.5$\,Hz and, correspondingly, discharge probabilities higher than ${\sim}3\times10^{-3}$.

The measurement of a single point on \figref{fig:sim:Multiplication} takes no longer than two hours in order to avoid large gain variations due to atmospheric conditions. It should be noted that the gain curve is obtained before each measurement session (defined by a gas mixture and$d_{\mathrm{source}}$). The latter is then accomplished within several hours after the gain is evaluated to assure the validity of the multiplication values throughout the complete session. 

The discharge probability depends strongly on the gas mixture and the differences are most prominent for the Ar-CO$_2$ and \NeCOtwo mixtures for which the amount of quencher is the same. For these two gases, the measurements have been carried out for four different values of $d_{\mathrm{source}}$. For a given gain, the discharge probability in \ArCOtwo is higher by several orders of magnitude than in \NeCOtwo. The difference can be explained by basic features of the corresponding noble gas. The range of alpha particles in the Ar-based mixture is almost \SI{40}{\percent} shorter compared to the Ne-based. Additionally, the number of primary electrons liberated by the incident particle is higher in Argon due to the lower value of the effective ionisation potential (see \Tabref{tab:sim:gas} for details). 
Accordingly, higher local charge densities are obtained in Ar-based gas mixtures. As a consequence, it is more likely to exceed the critical charge limit and hence the discharge probability may be effectively enhanced. 

For $d_{\mathrm{source}}=$ \SI{3.95}{\centi\meter} the discharge curves were measured for all four mixtures. Both Ar-based mixtures have lower breakdown limits than any of the Ne-based. Concerning the addition of a small fraction of N$_2$ to \NeCOtwo, the larger amount of quencher clearly enhances the stability of the mixture, while the transport properties remain almost unaffected, as discussed in \cite{GarabatosVCI}. A similar argument can be used to interpret the results with the \ArCOtwoThirty mixture. The larger amount of quencher increases the stability of the GEM, however, taking into account the alpha range and the effective ionisation potential of this gas, one would expect similar charge densities and thus discharge probability as in the 90-10 mixture. Therefore, the resulting difference can be assigned to the different transport properties in this mixture, which accordingly alter the density of primary ionisation.

\begin{figure*}[p]
\centering
\includegraphics[width=0.9\textwidth]{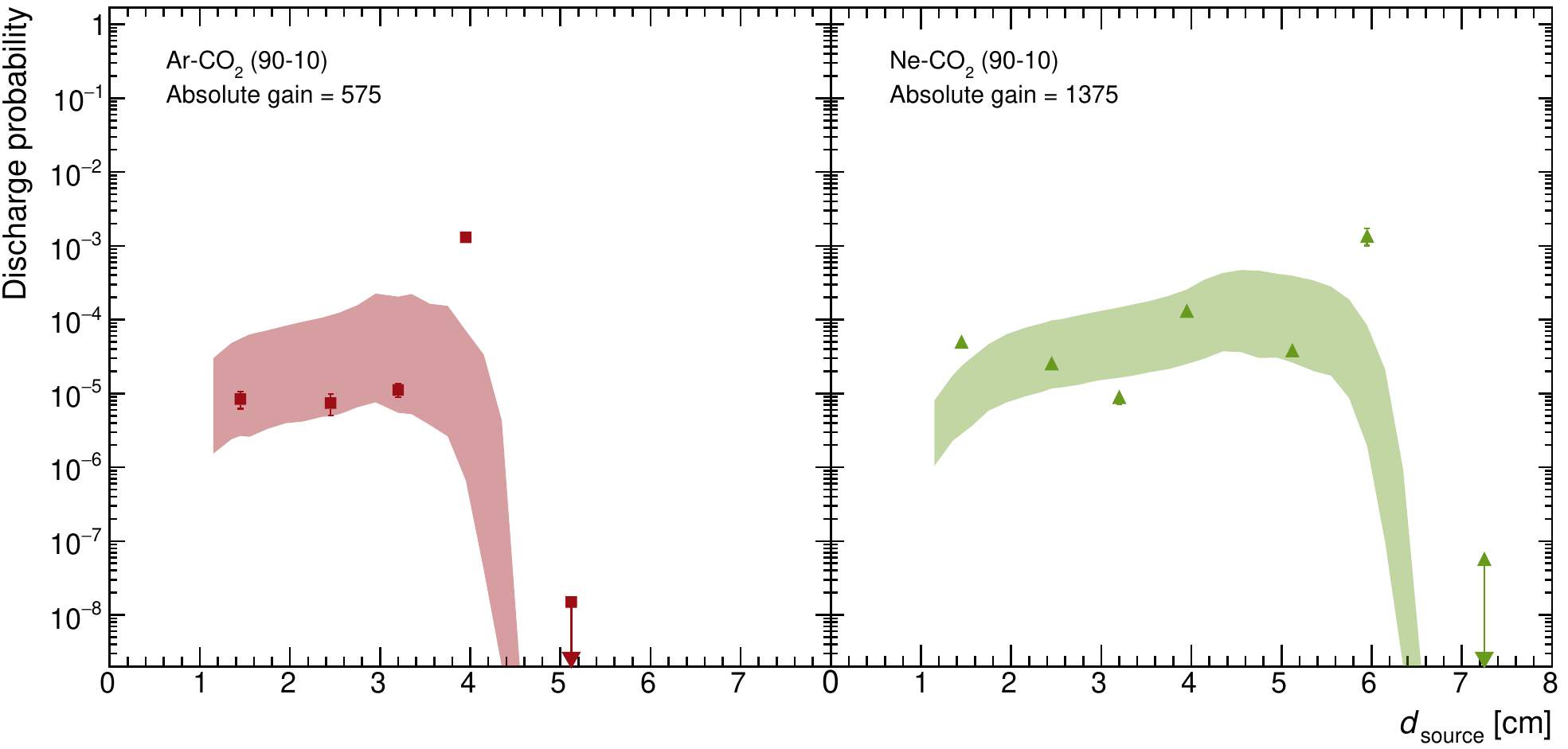}
\caption[Discharge probability as a function of the drift length at a fixed value of GEM absolute gain.]{(\textit{Colour online}) Discharge probability as a function of the drift length at a fixed value of GEM absolute gain. The bands indicate the outcome of the simulation, while the points correspond to measurements. The integration time in the simulation is \SI{50}{\nano\second} for \NeCOtwo, \SI{30}{\nano\second} for \ArCOtwo and \SI{40}{\nano\second} for \ArCOtwoThirty and \NeCOtwoNtwo.
The uncertainties of the measurement are typically smaller than the marker size, while the arrow for the last point indicates an upper limit. The width of the simulation bands is related to the range of the value of critical charge density. See text for details.}
\label{fig:sim:Range}
\end{figure*}

It is worth noting that there are no discharge curves measured in Ar-CO$_2$ (90-10) for higher values of $d_{\mathrm{source}}$. The reason for that is that the discharge probability for larger distances drops significantly by many orders of magnitude and consequently no discharges can be measured within a reasonable amount of time with a source rate of $\sim$\SI{600}{\hertz}. It is therefore relevant to look into the dependence on the distance between the source and the GEM foil in more details. 

\subsection{$d_{\mathrm{source}}$ dependence}
As there is no single gain value for which the discharge probability is measured at all distances, the experimental results presented in \figref{fig:sim:Multiplication} are fitted with an exponential function (not shown in the figure) to extra-/interpolate the discharge probability to the gain values of interest. The validity of this approach was tested by comparing the such obtained values to the measured data where possible. No significant differences were found within the uncertainties.

\Figref{fig:sim:Range} shows the discharge probability measured as a function of the distance $d_{\mathrm{source}}$ for a given absolute gain of 575 and 1375 for \ArCOtwo and \NeCOtwo, respectively. The measured dependencies indicate a monotonical increase of the discharge probability with $d_{\mathrm{source}}$. Beyond a certain distance ($\sim$\SI{5}{\centi\meter} and $\sim$\SI{7}{\centi\meter} for Ar- and Ne-CO$_2$, respectively) the discharge probability drops abruptly by several orders of magnitude such that only upper limits can be measured. The upper limit is defined as $1/\left(tR\right)$, where $t$ is the measurement time in which no discharge has been recorded. It should be noted, that the upper limits are measured for up to 30 hours at gains higher than indicated in \figref{fig:sim:Range} which allow us to neglect possible gain variations due to change of atmospheric conditions.

The distinctive behaviour of the measured dependence can be interpreted in the following way. An enhanced discharge probability is expected when alpha particles can pass through the GEM holes or are stopped in their closest vicinity, which in both cases yields the highest local primary charge densities in single holes.
When the distance is increased such that the alphas do no longer reach the GEM structure the discharge probability drops by orders of magnitude. Even though the primary ionisation still reaches the GEMs due to the charge transport in the drift field, the resulting charge density is then already significantly diluted by diffusion and will therefore have less probability to create a discharge. 

The local energy deposit of the alpha particle is significantly enhanced towards the end of its trajectory, which is manifested by an increase of the discharge probability  with increasing $d_{\mathrm{source}}$ values. In our case, however, this effect is convoluted with the emission of the particle within a solid angle element defined by the finite source dimensions and the hole in the cathode PCB (see \secref{sec:setup} for more details). Due to this complication, a simple comparison of the discharge curves in \figref{fig:sim:Range} to the Bragg curve is not possible. Therefore, we employ full-scale GEANT4 simulations of the setup, which allows us to explicitly compute the local energy deposit. In order to further verify the charge density hypothesis, we estimate the resulting charge density after the drift inside single GEM holes with the aim to obtain the critical charge limit responsible for the formation of discharges. It should be noted that similar studies for the Micromegas (Micro-Mesh Gaseous Structure) case has been performed in \cite{Procureur}. 
\section{Simulations}
\label{sec:sim}

\subsection{Detector simulation}
\label{sec:geant}
For the simulation of the energy deposit in the active detector medium, we use the latest version of GEANT4 (4.10.2) \cite{GEANT4}, which is a commonly used tool for the simulation of interactions of charged particles traversing matter in high-energy physics. The particle transport in GEANT4 is performed step-wise with an interaction taking place after each step. The distance between the steps is randomly sampled from the mean free path of the particle computed taking into account the cross sections of specified physics processes, which are summarised in a so-called physics list. Among the recommended physics lists for low-energetic applications \cite{PhysicsList}, we employ \texttt{G4\-Em\-Liver\-more\-Physics} in the course of our studies. 

The description of the setup follows closely the geometry of the detectors described in \secref{sec:setup}. A particle gun on the cathode randomly emits alpha particles with energies and intensities corresponding to the specifications of the mixed alpha-source as summarised in \Tabref{tab:Source}. The effect of the drift field in the active volume is included in the simulation, however without considering inhomogeneities resulting from the missing field cage. The resulting bias is found to be negligible. 
The exact position and energy deposit of each individual GEANT4 hit is then used in the further analysis steps. In total, about 500 million events are simulated and processed for each gas which is sufficient to compare the outcome of the simulation to the whole range of measured discharge probabilities. 

Of particular interest for the formation of discharges is the energy deposit of the alpha-particles in the closest vicinity of the GEM foil and the accordingly obtained charge densities inside single GEM holes. While the discharge itself is a fast phenomenon taking place on time scales $\mathcal{O}($\si{\nano\second}$)$, it is not known how long it takes to create the conditions in which a discharge can develop, i.e. how long the charges can accumulate inside the GEM hole before a discharge occurs. This time is linked to the exact mechanism underlying the discharge formation and could thus be used to differentiate between different possible scenarios, such e.g. the accumulation of ion space charge.
Therefore, we introduce the integration time \tInt as a parameter in our model.

\begin{figure}
\centering
\includegraphics[width=0.8\linewidth]{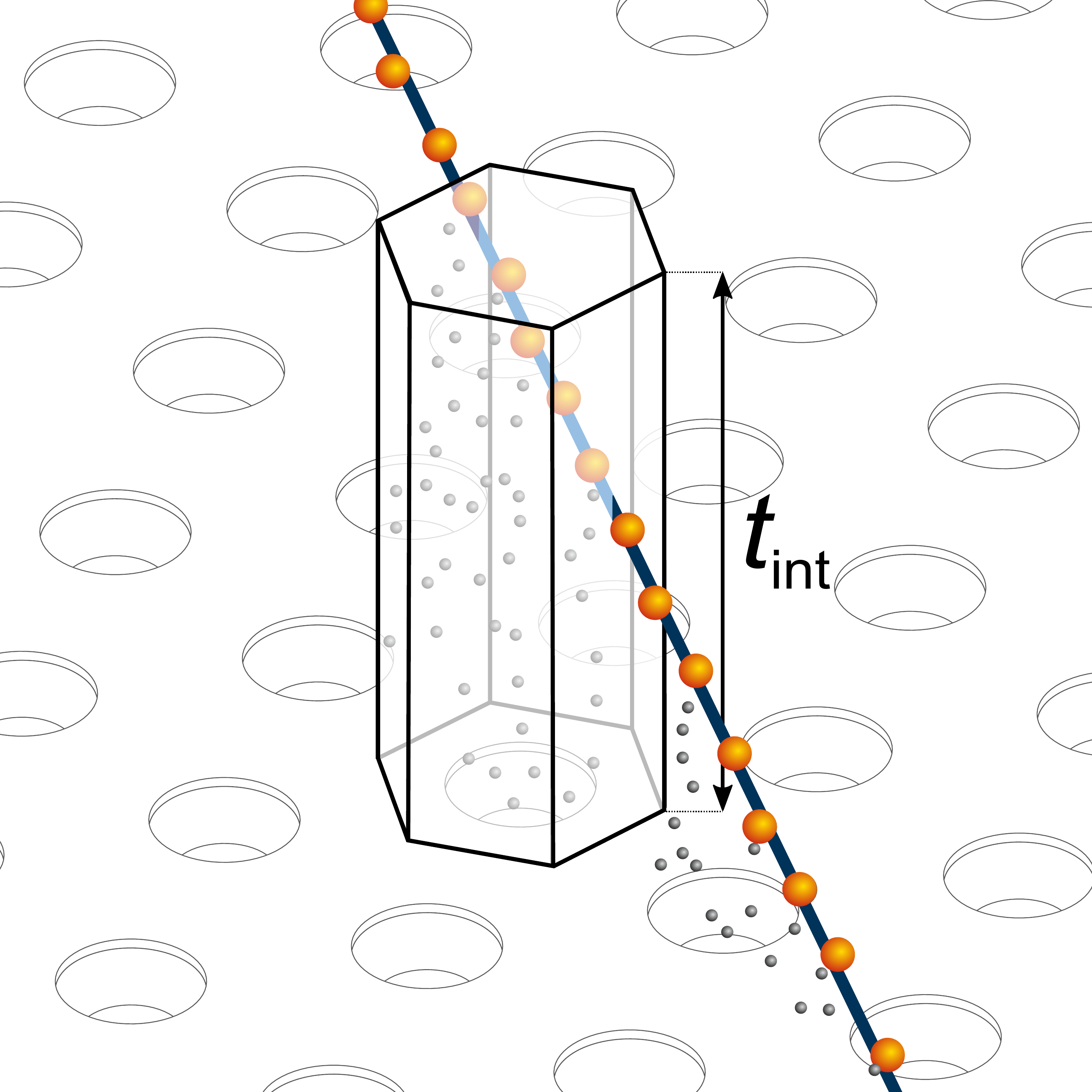}
\caption[Schematic sketch of honeycomb grid containing the area of interest around a single GEM hole.]{(\textit{Colour online}) Schematic sketch of honeycomb grid containing the area of interest around a single GEM hole. The incident alpha particle is represented by the blue line, whilst the GEANT hits are displayed as yellow balls. The resulting ionisation electrons are the gray dots. The height of the hexagonal shape defines the upper limit for charges that are taken into account for the accumulation inside the GEM hole at its bottom. See text for details.}
\label{fig:sim:GEMpattern}
\end{figure}

The number of ionisation electrons is obtained by dividing the energy deposit of each GEANT4 hit by the effective ionisation energy $W_{\mathrm{i}}$, which is depicted together with all other relevant gas parameters in \Tabref{tab:sim:gas}. All ionisation electrons located $d_{\mathrm{int}} = \tInt \cdot v_{\mathrm{drift}}$ above the GEM plane are then projected onto the latter. Transverse and longitudinal diffusion is taken into account by smearing the respective widths of the electron cloud by Gaussian distributions according to the values reported in \Tabref{tab:sim:gas}.
The electrons are then sorted into the honeycomb-like grid of the GEM foil, taking into account the collection efficiency of the GEM which is \SI{100}{\percent} for the drift field used. This procedure is illustrated in \figref{fig:sim:GEMpattern}.
The charges accumulated inside single GEM holes are then multiplied by the multiplication factor of the GEM. 
The single electron multiplication distribution in a single GEM follows a Polya distribution \cite{SingleElectron}, which accommodates the fluctuations of the amplification process. 
In our case, however, a very large number of electrons is accumulated inside a single GEM hole and therefore, according to the central limit theorem, the shape of the fluctuations is expected to become Gaussian and hence symmetric. Accordingly, on average no impact on the discharge probability is expected.
Therefore, fluctuations of the avalanche are not considered in the model, which allows us to investigate the impact of solely the primary charge density on the discharge probability.

As the dependence of the discharge probability on the accumulated and multiplied charges inside a GEM hole is unknown, we are left with the rather crude assumption of a fixed threshold of accumulated charges \Qcrit. Accordingly, a discharge in the simulation is defined as an event in which the number of accumulated charges inside a single GEM hole exceeds the value of the critical charge limit \Qcrit.
This corresponds to a situation similar to the well-known Raether limit \cite{Raether} for parallel plate detectors. 
The exact critical limit valid for GEM-based detectors will be estimated in the following way. The final discharge probability is defined as the number of events in which \Qcrit is exceeded normalised to the number of simulated events for a given drift gap and GEM multiplication factor and remarkably requires no further normalisation. Note that only one such event per drift gap, multiplication and event is possible as in this case the potential difference across the GEM rapidly decreases and thus prevents the formation of further discharges caused by the very same incident particle.

The simulated energy deposit of the alpha-particles varies with the penetration depth, reaching a distinctive maximum at the very end of the particle's track (Bragg peak). Therefore, a significant dependence of the GEANT4 step length on the latter is expected. Consequently, in particular for very small integration times, the step length could be larger than the actual integration volume and accordingly some hits might not be taken into account in the analysis. An additional impact on the simulated energy deposit may be introduced by the choice of physics list, which steers the energy loss of the incident particle. By varying the step length between 1 and \SI{1000}{\micro\meter} and employing a different physics list (\texttt{G4\-Em\-Standard\-Physics\_\-option4}), we find that the results are robust against the choice of the latter and only impacted by very large step lengths $\mathcal{O}$(\SI{100}{\micro\meter}). Hence, we fix the step length to \SI{10}{\micro\meter} and smear the distance between the GEANT4 hits according to a flat distribution.

By not considering the field distortions resulting from the missing field cage, an additional bias may be introduced to the electron transport in the drift field. Distortions of the drift field result in a non-constant drift velocity both with respect to magnitude and direction, which would demand for a more sophisticated treatment of the constant integration time \tInt.
In order to estimate the impact of the above described effect, we have conducted finite element calculations using the the COMSOL Multiphysics$\textsuperscript{\textregistered}$ software \cite{COMSOL} and found variations of the drift field below \SI{1}{\percent} for the integration times of our interest, which are therefore neglected.

\subsection{Comparison to experimental data}
In order to compare the measurements presented in the previous section to the outcome of the simulation, we vary $d_{\mathrm{source}}$ in steps of \SI{2}{\milli\meter} and the multiplication factor of the GEM in steps of 50 for multiplication factors smaller than 1500 and in steps of 250 above that.

In order to study the dependence of the two free parameters of the model, \tInt and \Qcrit, we compare the various simulation outcomes with the data varying \tInt between 2 and 400\,ns and \Qcrit in steps of \num{5e4} electrons between \num{1e5} and \num{1.25e7}. The best values of both parameters are then found with the help of a \chiSquared minimisation procedure.

\begin{figure}
\centering
\includegraphics[width=\linewidth]{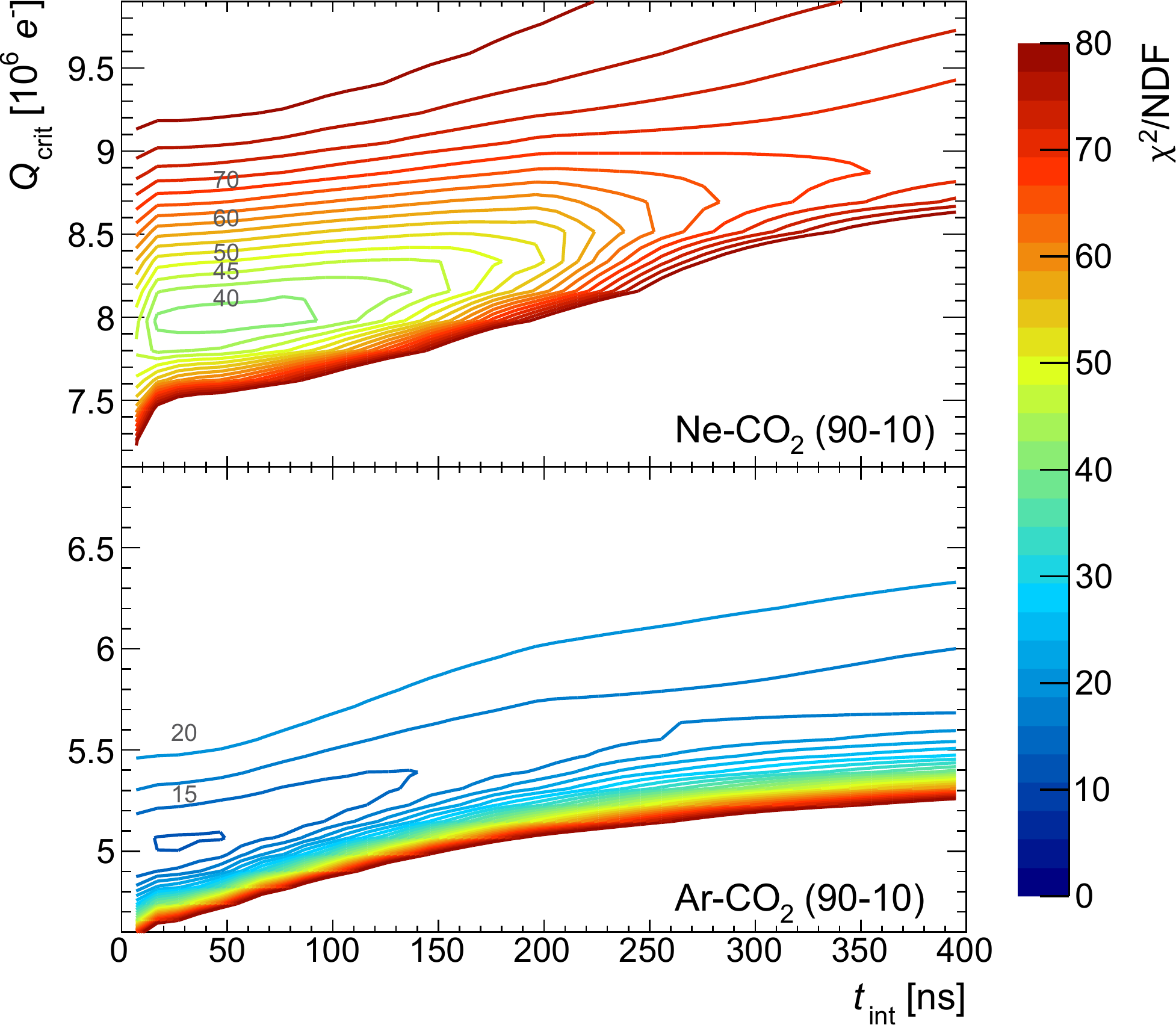}
\caption[The dependence of the value of \Qcrit on the integration time \tInt. The contours represent the reduced \chiSquared of the model comparison to the discharge probability]{(\textit{Colour online}) The dependence of the value of \Qcrit on the integration time \tInt for Ar-CO$_2$ and \NeCOtwo. The contours represent the reduced \chiSquared of the model comparison to the discharge probability for a given \tInt and \Qcrit. See text for details.}
\label{fig:sim:Chi2Range}
\end{figure}

According to our observation, the data displayed in \figref{fig:sim:Range} are most sensitive to variations of \tInt. Therefore, we compare the simulation outcomes to the measurements presented in this plot to constrain the value of this parameter. Both \tInt and \Qcrit are varied within the above specified range for each gas independently.
For each combination of these values the goodness of fit \chiSquared is computed by comparing to the experimental data. 
The lowest \chiSquared determines the best value of \tInt with the corresponding \Qcrit as displayed in \figref{fig:sim:Chi2Range} for Ar-CO$_2$ and \NeCOtwo.
For \NeCOtwo, the lowest values of \chiSquared are achieved for integration times ranging between 20 and 90\,ns, while for \ArCOtwo, the minimum is around 15 to 50\,ns. The resulting critical charge limit \Qcrit, however, varies by about \SI{10}{\percent} over the whole range of the probed values of $t_{\mathrm{int}}$ as shown in \figref{fig:sim:Chi2Range}.

The best value of \tInt is then \SI{50}{\nano\second} for \NeCOtwo and \SI{30}{\nano\second} for \ArCOtwo. 
If additionally all data of \figref{fig:sim:Multiplication} are considered for the best value of \tInt, the \chiSquared optimisation yields slightly different values of \Qcrit for different graphs. The resulting variation is below \SI{10}{\percent} and accommodated by the bands shown in Figs. \ref{fig:sim:Multiplication} and \ref{fig:sim:Range} around the mean value of \Qcrit for each gas with a width is defined by its RMS. The resulting values of \Qcrit for Ar-CO$_2$ and Ne-CO$_2$ (90-10) mixtures are summarised in \Tabref{tab:sim:Qcrit}.

As only one $d_{\mathrm{source}}$, i.e. one graph in \figref{fig:sim:Multiplication}, can be used to constrain \Qcrit for \ArCOtwoThirty and \NeCOtwoNtwo, it is not possible to apply the full \chiSquared minimization technique to find a best value of \tInt. Therefore, an integration time of \SI{40}{\nano\second} is chosen which is a good compromise of the above found values of \tInt. Even though the variation of \Qcrit within the range of \tInt is negligible, the resulting values should be considered with due diligence and are therefore omitted in \figref{fig:sim:Chi2Range}. The values of \Qcrit for \ArCOtwoThirty and \NeCOtwoNtwo are of the order of $5 \times 10^{6}$ and $9 \times 10^{6}$, respectively, and therefore compatible with the results in \Tabref{tab:sim:Qcrit}. 

\begin{table}
\centering
\caption{Values of critical charge \Qcrit for different gas mixtures and an integration time \tInt of \SI{50}{\nano\second} for \NeCOtwo and \SI{30}{\nano\second} for \ArCOtwo, respectively. }
\begin{tabular}{l c }
\toprule
\textbf{Gas} & \textbf{\textit{Q}}$_{\mathrm{\textbf{crit}}}$ \\
\midrule
\ArCOtwo & $(4.7 \pm 0.6) \times 10^{6}$ \\ 
\NeCOtwo & $(7.3 \pm 0.9) \times 10^{6}$ \\
\bottomrule
\end{tabular}
\label{tab:sim:Qcrit}
\end{table}

\subsection{Conclusions from the model}
The comparison of the model, taking into account only the primary charge density at the GEM plane, to the measurements in \figref{fig:sim:Multiplication} and \figref{fig:sim:Range} yields reasonable agreement and demonstrates the validity of the assumption of a critical charge limit underlying the discharge formation. Remarkably, the model requires no further normalisation to fit to the data. 
The resulting values for this critical charge limit \Qcrit are in line with the predictions based on the Raether limit and the values presented in \cite{Bressan} and differs between the gas mixtures used in the detector.

In our model we have introduced the integration time \tInt to mimic the impact of charge accumulation inside the GEM holes. 
For integration times larger than the transfer time of electrons $\mathcal{O}($\si{\nano\second}$)$ \cite{GEMgain}, clearly an influence of a mechanism different than the simple accumulation of electrons is expected. 
This holds, in particular, to the case in which charge accumulation prior to a discharge occurs over a time of a few tens of ns, which is of the same order as the transfer time of the ions produced in the avalanche to the top GEM electrode \cite{GEMgain}. 
The accordingly accumulating space charge modifies the amplification field and thus leads to the development of a streamer, as also concluded in \cite{Resnati}. Our results on the integration time \tInt support this hypothesis. 

The difference of the values of \Qcrit and \tInt for the different gases might be related to the precise mechanism how the discharge forms in the respective medium.

\section{Summary}
\label{sec:summary}
We have performed systematic studies of the discharge probability in a single GEM detector exposed to heavily ionising alpha particles in Ar- and Ne-based gas mixtures. Our experimental findings imply that large clusters of primary charge are the main reason for the discharge formation in a GEM detector. We show that the occurrence of sparks is more likely in argon than in neon mixtures since in the latter the enhanced range of alpha particles and higher ionisation potentials result in lower primary charge densities. 

The hypothesis was confirmed by Monte-Carlo simulations in which we explicitly calculate the number of primary electrons liberated by alpha particles at a given distance between the source and the GEM.
Taking into account the primary ionisation distribution and basic transport properties of the gas mixtures, we are able reproduce the experimentally measured discharge probability by normalising the number of events in which the charge contained inside a single GEM hole exceeds a critical charge threshold $Q_{\mathrm{crit}}$ to the total number of emitted alpha particles.

Simulation and experimental data sets allowed us to extract the critical charge density which leads to a discharge in a GEM hole. We obtain $Q_{\mathrm{crit}}$ of $\left(4.7\pm0.6\right)\times10^6$ and  $\left(7.3\pm0.9\right)\times10^6$ electrons per GEM hole for Ar-CO$_2$ (90-10) and Ne-CO$_2$ (90-10) mixtures respectively. The difference may point to an additional influence of gas properties on the discharge formation process. 

We conclude that it is the number of particles that cross the GEM volume which determines the discharge rate of the detector, and that the primary charge that reaches the readout chambers after a (relatively) long drift has significantly less impact on the detector stability. 
 
In multi-GEM systems the total gain can be shared between several foils building a stack, and thus increasing its stability against electrical discharges. However, the local charge densities approaching a single hole in any GEM of the stack should be regarded as the driving criterion to trigger sparks. 
With the expected primary charge distribution at hand, the approach presented in this paper allows one to estimate the spark rate in future experiments.

\newenvironment{acknowledgement}{\relax}{\relax}
\begin{acknowledgement}
\section*{Acknowledgements}
The Authors wish to thank C. Garabatos (GSI Darmstadt) for providing the Magboltz calculations and careful proofreading of this manuscript and B. Ketzer (Bonn University) for his valuable help in the starting phase of the project. 

This work was supported by the DFG Cluster of Excellence "Origin and Structure of the Universe" (www.universe-cluster.de) [project number DFG EXC153]; by the Federal Ministry of Education and Research (BMBF, Germany) [grant number 05P15WOCA1].

Computations for the project (t2321) were made on the high performance computer Linux Cluster at the Leibniz Supercomputing Centre (LRZ) of the Bavarian Academy of Sciences and Humanities. The operation of this supercomputer is funded via the Bavarian State Ministry of Education, Science and the Arts.
\end{acknowledgement}





\section*{References}
\bibliographystyle{elsarticle-harv} 
%

\end{document}